\documentclass[aps,twocolumn,prb,floatfix, superscriptaddress]{revtex4-2}
\usepackage[T1]{fontenc}
\usepackage{mathptmx} 
\usepackage[english]{babel}
\usepackage[utf8]{inputenc}
\usepackage{polski}
\usepackage{indentfirst}
\usepackage{amsthm}
\usepackage{amsmath}
\DeclareMathOperator{\sinc}{sinc}
\usepackage{lmodern} 
\usepackage{graphicx}
\usepackage{geometry}
\usepackage[caption=false]{subfig}
\usepackage{hyperref}
\usepackage{relsize}
\usepackage[capitalize]{cleveref}
\newgeometry{tmargin=1.5cm, bmargin=1.5cm, lmargin=1.5cm, rmargin=1.5cm}

\begin{document}

\title{Spin-wave Talbot effect in thin ferromagnetic film}
\author{Mateusz Gołębiewski} \email{matgol2@amu.edu.pl}
\affiliation{Faculty of Physics, Adam Mickiewicz University, Uniwersytetu Poznańskiego 2, 61-614 Poznań, Poland}
\author{Paweł Gruszecki}
\affiliation{Faculty of Physics, Adam Mickiewicz University, Uniwersytetu Poznańskiego 2, 61-614 Poznań, Poland}
\affiliation{Institute of Molecular Physics, Polish Academy of Sciences, Mariana Smoluchowskiego 17, 60-179 Poznań, Poland}
\author{Maciej Krawczyk} \email{krawczyk@amu.edu.pl}
\affiliation{Faculty of Physics, Adam Mickiewicz University, Uniwersytetu Poznańskiego 2, 61-614 Poznań, Poland}
\author{Andriy E.~Serebryannikov}
\affiliation{Faculty of Physics, Adam Mickiewicz University, Uniwersytetu Poznańskiego 2, 61-614 Poznań, Poland}

\begin{abstract}
The Talbot effect has been known in linear optics since XIX century and found various technological applications. In this paper, with the help of micromagnetic simulations, we demonstrate the self-imaging phenomenon for spin waves in a thin, out-of-plane and in-plane magnetized ferromagnetic film, whose propagation is described by Landau–Lifshitz nonlinear equation. We show that the main features of the obtained Talbot carpets for spin waves can be described, to a large extent, by the approximate analytical formulas yielded by the general analysis of the wave phenomena. Our results indicate a route to a feasible experimental realisation of the Talbot effect at low and high frequencies and offer interesting effects and possible applications in magnonics. 
\end{abstract}

\date{\today}

\maketitle

\section{Introduction}
Spin waves (SWs) are coherent disturbances of the magnetization which may propagate in magnetic material as waves. In ferromagnetic materials, the SW dynamics is determined by strong isotropic exchange interactions coexisting with anisotropic magnetostatic interactions. In thin ferromagnetic films, the magnetostatic interactions cause the SW properties being strongly dependent on the magnetization orientation with respect to the film plane, and also dependent on the relative orientation of the propagation direction and the static magnetization vector. This makes the studies of SWs interesting and offers properties uncommon for other types of waves, like negative group velocity, caustics, and dynamic reconfigurability control. Their frequency spans range from a few to hundreds GHz with the corresponding wavelengths range extended from micrometers to tens of nanometers, which make them very attractive for applications in microwave technology. In this context, it is interesting to test in magnonics the basic laws that govern waves phenomena and search for analogs of the effects known for electromagnetic or acoustic waves. Basic equations describing propagation of SWs differ from the ones for electromagnetic and acoustic waves, so that justification of an analogue of each phenomenon known for the latter invokes solution of Landau–Lifshitz equation for the former -- this approach was used in demonstration of the SW graded index lenses \cite{graded_index_lenses}, SW Luneburg lenses \cite{Luneberg} and SW Fourier optics \cite{SW_Fourier}, to name a few. Recently, the Snell's law for SWs in thin ferromagnetic films \cite{Sti16}, mirage effect \cite{Gru18}, and spin-wave Goos-H\"anchen effect have been predicted and demonstrated \cite{Gru17, stigloher2018observation}. Also, self-focusing of SWs \cite{Dem08}, SW diffraction on gratings \cite{Man12}, and formation of the SW beams \cite{Kho04, gieniusz2017switching, korner2017excitation} found experimental confirmation. The analogs of the graded refractive index structures \cite{graded_index_lenses,Dav15,vogel2018control}, metamaterials \cite{Mikhaylovskiy2010,Mruczkiewicz2012}, and metasurfaces \cite{Zel19}, have recently been introduced to magnonics. Thus, it may be expected that other phenomena and concepts are also transferable to magnonics, but these need to be verified. Among them, the Talbot effect (self-imaging effect) should be mentioned.

The Talbot effect was observed for light in XIX century \cite{Talbot36} and then explained in Ref.~[\onlinecite{Rayleigh81}]. In recent years, this effect has been extensively revisited, \textit{e.g.}, see \cite{Wen13} and references therein. It has been used to improve x-ray imaging \cite{Bravin_2012} and advance the process of lithographic patterning \cite{Sat14,Zho16,Vet18}, and proposed for realization of some physical models and computing scenarios \cite{Big08,Far15,Saw18}, the applications of which can be interesting also for magnonics. Apart from electromagnetic waves propagating in a medium, the Talbot effect has already been demonstrated for plasmons \cite{Den07}, waves in fluids \cite{Sun18, Bakman19}, and exciton-polaritons \cite{Gao16}, but for SWs it has not been shown so far.

In this paper, we demonstrate the Talbot effect with the use of micromagnetic simulations for SWs propagating in a thin ferromagnetic film magnetized out-of-plane. We show, that the diffraction grating created by the periodically located holes in a thin ferromagnetic film allows obtaining Talbot carpets, which are formed by the SWs propagating in the film and passing through the grating. The demonstration is presented at high frequencies where the exchange interactions dominate, and at low frequencies where the magnetostatic interactions contribute, too. Furthermore, we study also the influence of SW damping on the Talbot effect to show possibility for its experimental verification and we perform simulations of the Talbot effect for in-plane magnetized permalloy (Py) film in order to check the impact of changing the external magnetic field direction on the Talbot carpets and the possibility of using smaller field values.

The structure of the paper is as follows. In Sec.~\ref{Sec:model}, we present derivation of the main parameters of the Talbot pattern based on the wave optics model. Then, we focus on the case of out-of-plane magnetization. In Sec.~\ref{Sec:Mic}, we present the results of micromagnetic simulations for short and long SWs, in Sec.~\ref{Sec:exp_feas} we study the impact of SW damping on potential applications and in Sec.~\ref{Sec:in_plane} we carry out simulations for in-plane magnetization. Finally, we summarize our results in Sec.~\ref{Sec:conclusions}.
The appendices are put at the end, which present the model development and analysis of some issues presented in the main part of the paper.

\section{Talbot effect -- model description }
\label{Sec:model}
\subsection{Talbot length}
\label{Sec:len}
The phenomenon of self-imaging, known as the Talbot effect, results from the interference of plane wave passed through an array of the periodically arranged objects, which often represents a diffraction grating. Using the knowledge that SWs may behave similarly to electromagnetic and acoustic waves, we can explain the essence of the SW Talbot effect with the help of an analysis of the basic wave phenomena, such as diffraction and interference. According to the principle formulated by Christian Huygens \cite{huygens} and later supplemented by Augustin Fresnel \cite{fresnel}:
\begin{quote}
Each unobstructed point of the wavefront at a given moment acts as a source of secondary spherical elementary waves with the same frequency as the primary wave. The amplitude of the resultant field at any other point is the superposition of all these elementary waves taking into account their amplitudes and relative phase differences.
\end{quote}
This principle is particularly important not only for understanding the phenomenon of wave diffraction, but first of all, for the design of the simulation system and building a suitable mathematical model. It can be used when an aperture width is comparable to a length of incident plane wave, an angle of secondary waves is so large that this aperture can be treated as an elementary source of circular waves (or cylindrical in a two-dimensional view), and the entire diffraction grating can be treated as a one-dimensional infinite matrix of the periodically spaced elementary sources of circular waves.
The presented concept of diffraction grating allows us to describe its effects and thus the diffraction field with a high accuracy, by means of the superposition principle.

Repetitive modulation of intensity occurring along the propagation direction of the waves diffracted on periodic obstacles, \textit{i.e.}, the Talbot self-imaging effect we study, defines the length parameter also called from his name. The Talbot length determines the period over which secondary beams re-focus their source image, \textit{i.e.}, when the diffraction image is created after passing through a diffraction grating, at a certain, precisely determined distance  $z_T$ in the direction perpendicular to the grating plane, an image identical to our original grating is formed. Classical formula for the Talbot length \cite{Rayleigh81} can be derived based on the geometrical consideration, which is known from wave optics. It yields  
\begin{equation}
z_T=m\frac{d^2}{\lambda}, 
\label{Eq:zt}
\end{equation}
where $m$ is an integer specifying the number of subsequent self-images, $d$ is a diffraction grating period, and $\lambda$ is an incident wave length. For even $m$ values, we obtain the distance between the primary (basic), not laterally shifted in phase Talbot images, whereas for odd $m$ values -- we observe the secondary Talbot images laterally shifted in phase by half the period of the diffraction grating, as shown in Fig.~\ref{Fig:Sch_T}. More details are given in Appendixes \ref{Sec:App_A} and \ref{Sec:App_B}.

\begin{figure}[htp]
\begin{center}
\includegraphics[width=9cm]{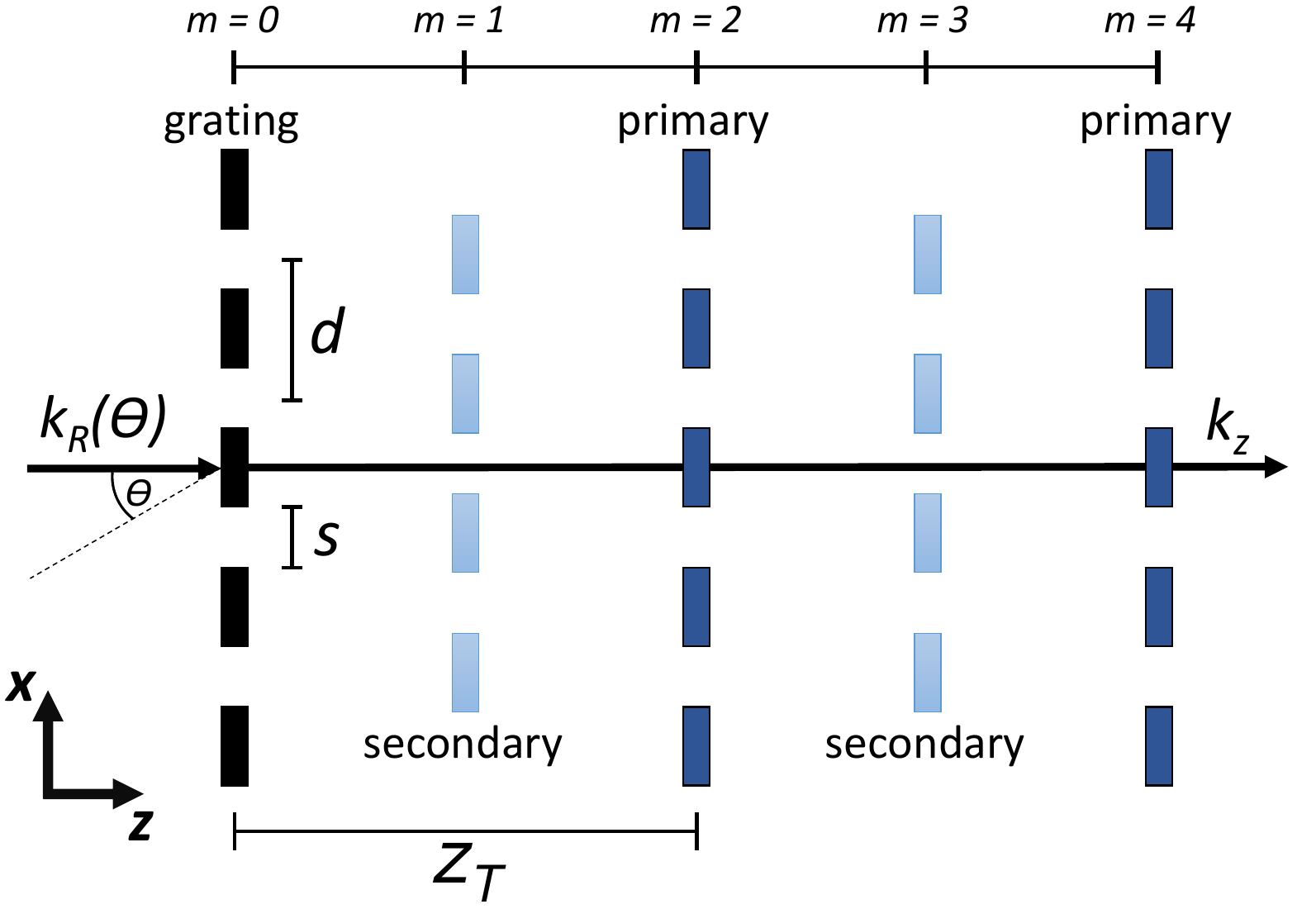}
\end{center}
\vspace{-0.6cm}
\caption{Diagram of self-images arising behind the diffraction grating with indicated Talbot length $z_T$, diffraction grating constant $d$, aperture width $s$, wavevector $k_R$ as a function of incident angle $\theta$, and wavevector $k_z$ propagating along the $z$-axis. \label{Fig:Sch_T}}
\end{figure}

\subsection{Talbot carpet}
\label{Sec:TalC}

In order to obtain an analytical expression for the intensity distribution of the waves after passing through a periodic, infinitesimally thin diffraction grating, we start with writing \cite{talbot-lau} the transformation equation at the grating in $z = 0$:
\begin{equation}
\psi(x,z=+0)=\psi(x,z=-0)\cdot t(x),
\label{Eq:psip}
\end{equation}
where $\psi(x,z)$ represents incident wave and $t(x)$ is a periodic grating transmission function. The term $z=-0$ means the coordinate $z=0$ at the incidence side of the diffraction grating, and $z=+0$ similarly means the coordinate $z=0$ at the transmission side of the grating. Under assumptions that the one-dimensional periodic structure has an infinite length along the $x$-axis and the incident beam is a plane wave propagating along the normal towards the grating, the transmittance of the diffraction grating, $t(x)$, can be defined as
\begin{equation}
t(x)=\sum_{n=-\infty}^{\infty} A_{n}\exp{\left[\frac{2{\pi}inx}{d}\right]}.
\label{Eq:tx}
\end{equation}
Formula (\ref{Eq:tx}) is so-called rectangular function, which means it has a ,,binary'' nature, \textit{i.e.}, a plane wave passes through the structure, when its transparent parts are illuminated, with the maximum value ($t(x)=1$), and do not pass through it when its opaque parts are illuminated ($t(x) = 0$). For the unlimited ($\mbox{max}|n|=\infty$) rectangular function along the $x$-axis, components of the Fourier decomposition $A_n$ are described by the formulas:
\begin{equation}
A_n=\sinc{(\frac{n\pi s}{d})},
\end{equation}
where $\mbox{sinc}(w)=\mbox{sin}w/w$, and 
\begin{equation}
A_0=s/d.
\end{equation}
Substituting Eq.~(\ref{Eq:tx}) to Eq.~(\ref{Eq:psip}), using a plane wave equation as $\psi(x,z=-0)$, and remembering that instead of the wavevector $k=2\pi/\lambda$, its projection on the $x$-axis depending on the incidence angle $\theta$, $k_R=k\sin{\theta}$, should be taken into account, we get
\begin{equation}
\psi(x,z=+0)=\sum_{n=-\infty}^{\infty}\!\!\!\!\!\!A_{n}\exp{[ix(k_dn+k_R)]},
\label{Eq:psi02}
\end{equation} 
where $k_d = 2\pi /d$.\\
Next, we extend the function $\psi$ to describe the wave propagation in the direction of positive $z$ values by 
\begin{equation}
\psi(z)=\exp{(ik_zz)},
\label{Eq:psiz}
\end{equation}
where $k_z$ is the axial wavevector associated with propagation along the $z$-axis.\\
Therefore, the complete expression, describing the plane wave after passing through the diffraction grating at arbitrary distance from it, can be written as
\begin{equation}
\psi(x,z)=\sum_{n=-\infty}^{\infty}\!\!\!\!A_{n}\exp{[ix(k_dn+k_R)+ik_zz]}.
\label{Eq:psixz}
\end{equation}
The axial wave number $k_z$ can be expressed as the function of the wavevector $k$ as follows: 
\begin{equation}
k^2=k_z^2+(k_dn+k_R)^2 \;\; \rightarrow \;\; k_z=\sqrt{k^2-(k_dn+k_R)^2}.
\label{Eq:kz}
\end{equation}
Assuming that $k \gg (k_d n + k_R)$ and using the Taylor's expansion so that
\begin{equation}
k\sqrt{1-\left(\frac{k_dn+k_R}{k}\right)^2} \;\approx\; k\left(1-\frac{1}{2}\left(\frac{k_dn+k_R}{k}\right)^2\right),
\label{Eq:kztayl}
\end{equation}
we finally obtain 
\begin{equation}
k_z \approx k-\frac{(k_dn+k_R)^2}{2k}.
\label{Eq:kz1}
\end{equation}
For the purposes of numerical study, we assume that the initial plane wave incident along the normal to the diffraction grating plane ($\theta = 0$). Then, we obtain the following formula:
\begin{equation}
\psi(x,z)\approx\sum_{n=-\infty}^{\infty}\!\!\!\!\!\!A_{n}\exp{\left[ixn\frac{2\pi}{d}+iz\left(\frac{2\pi}{\lambda}-\frac{n^2\lambda}{d^2}\right)\right]}.
\label{Eq:psixz0}
\end{equation} 
Here, the last term contains the inverse Talbot length formula, see Eq.~(\ref{Eq:zt}) for $m = 1$.\\
To generate energy density graphs in the near diffraction field, we calculate the field intensity function, \textit{i.e.}, 
\begin{equation}
I(x,z)=\psi(x,z)\cdot\psi^{*}(x,z).
\label{Eq:Ipsi}
\end{equation}

\begin{figure}[htp]
\begin{center}
\includegraphics[width=\linewidth]{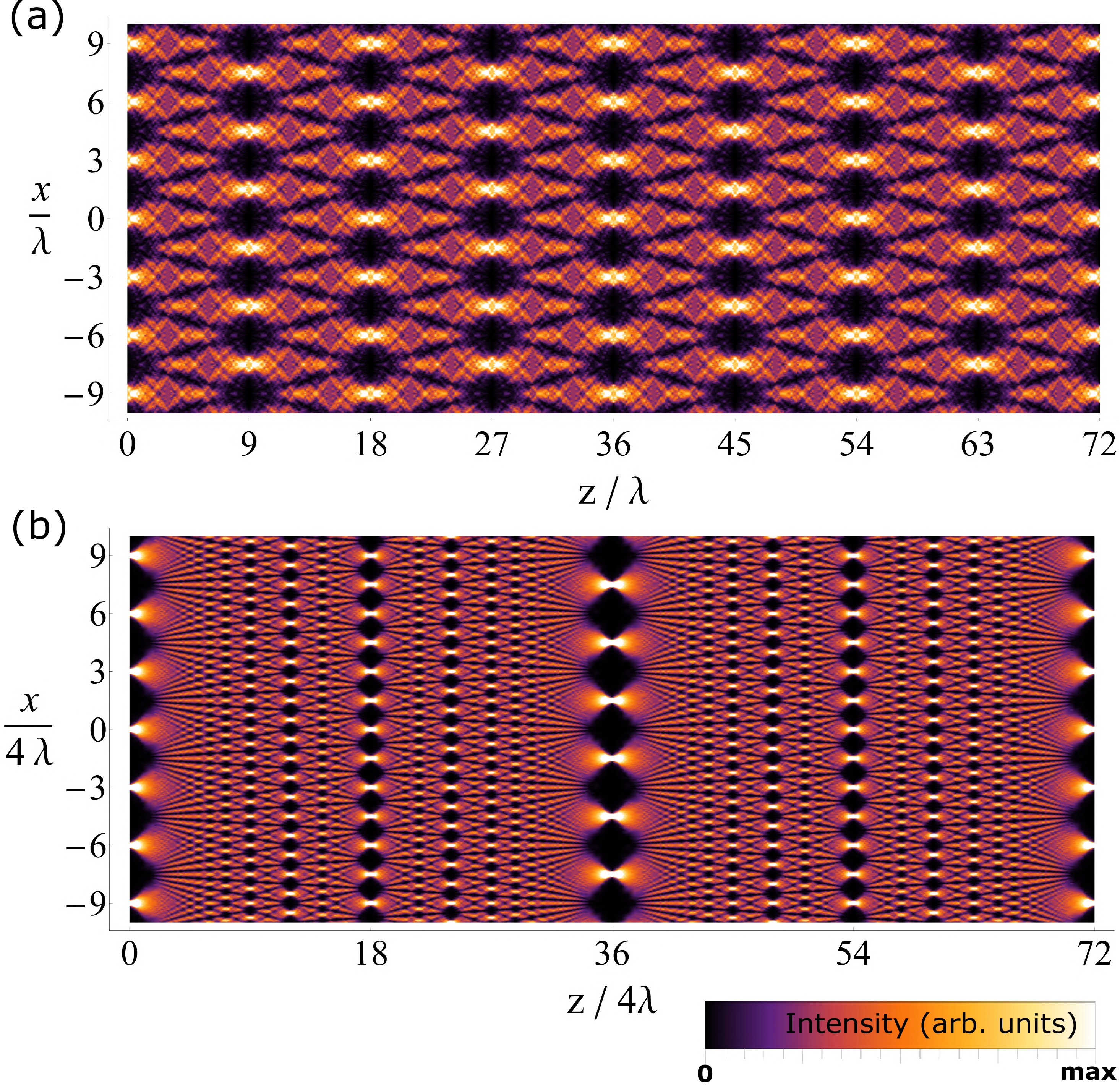}
\end{center}
\caption{Talbot carpets obtained from Eqs.~(\ref{Eq:psixz0}) and (\ref{Eq:Ipsi}) for 31 sum terms and: (a) the aperture width $s=\lambda=1 \; [\text{arb. units}]$ and diffraction grating constant $d=3\lambda$, and (b) $s=\lambda=0.25 \; [\text{arb. units}]$, $d=12\lambda$. The grating is located at the left edge of each plot.}
\label{Fig:s1d3}
\end{figure}

The theoretical results are shown in Fig.~\ref{Fig:s1d3} for selected parameters set and 31 sum components, in order to demonstrate the basic features of these Talbot carpets. They clearly show the primary and secondary Talbot images, as well as, the lower orders of these images (visible especially in Figure \ref{Fig:s1d3}b), which testify to the fractal nature of the effect \cite{fractal}. Being obtained with the use of rather general formalism, these results form the basis for understanding of and comparing with the numerical results presented for SWs in the next section.

\section{Micromagnetic simulations}
\label{Sec:Mic}

To demonstrate the Talbot effect for SWs and examine the theoretical assumptions and predictions of Sec.~\ref{Sec:TalC}, we have conducted a series of micromagnetic simulations by taking into account dipolar interactions and  employing MuMax3 software \cite{vansteenkiste2014design}. The simulations have been performed for an uniformly out-of-plane magnetized 5-nm-thick permalloy (Py) film, which is characterized by the following magnetic parameters: saturation magnetization $ M_\mathrm{S} = 860 $ kA/m, exchange stiffness $ A_\mathrm{ex} = 13$ pJ/m, gyrometric ratio $ \gamma = 176 $ rad GHz/T, and damping constant $ \alpha = 0.0001$. The film has been uniformly magnetized by an external magnetic field, $\mu_0 H_0 = 1.1 $ T ($H_0>M_\mathrm{S}$), which is directed perpendicular to the film's plane. The above given magnetic parameters are kept the same for all performed numerical simulations. We compute the diffraction of normally incident plane SWs of frequencies 3 GHz and 40 GHz for various parameter sets of a diffraction grating. The distance between individual holes creating the diffraction grating is adjusted  to approximately correspond to the incident SW wavelength (it is calculated analytically in Appendix~\ref{Sec:App_dispersion}). The simulated steady-state, \textit{i.e.}, the state with a fully evolved interference pattern, is analyzed. For better visualization of the Talbot effect, the periodic boundary conditions have been also used along the $z$-axis (\textit{i.e.}, perpendicular to the grating plane), at the edges of the simulated area, so that the diffraction grating was much longer compared to the near diffraction field range. Further details of simulations can be found in the Appendix~\ref{Sec:App_Simulation}.

\subsection{Spin-wave Talbot carpets and Talbot length}

\begin{figure}[htp]
\begin{center}
\includegraphics[width=\linewidth]{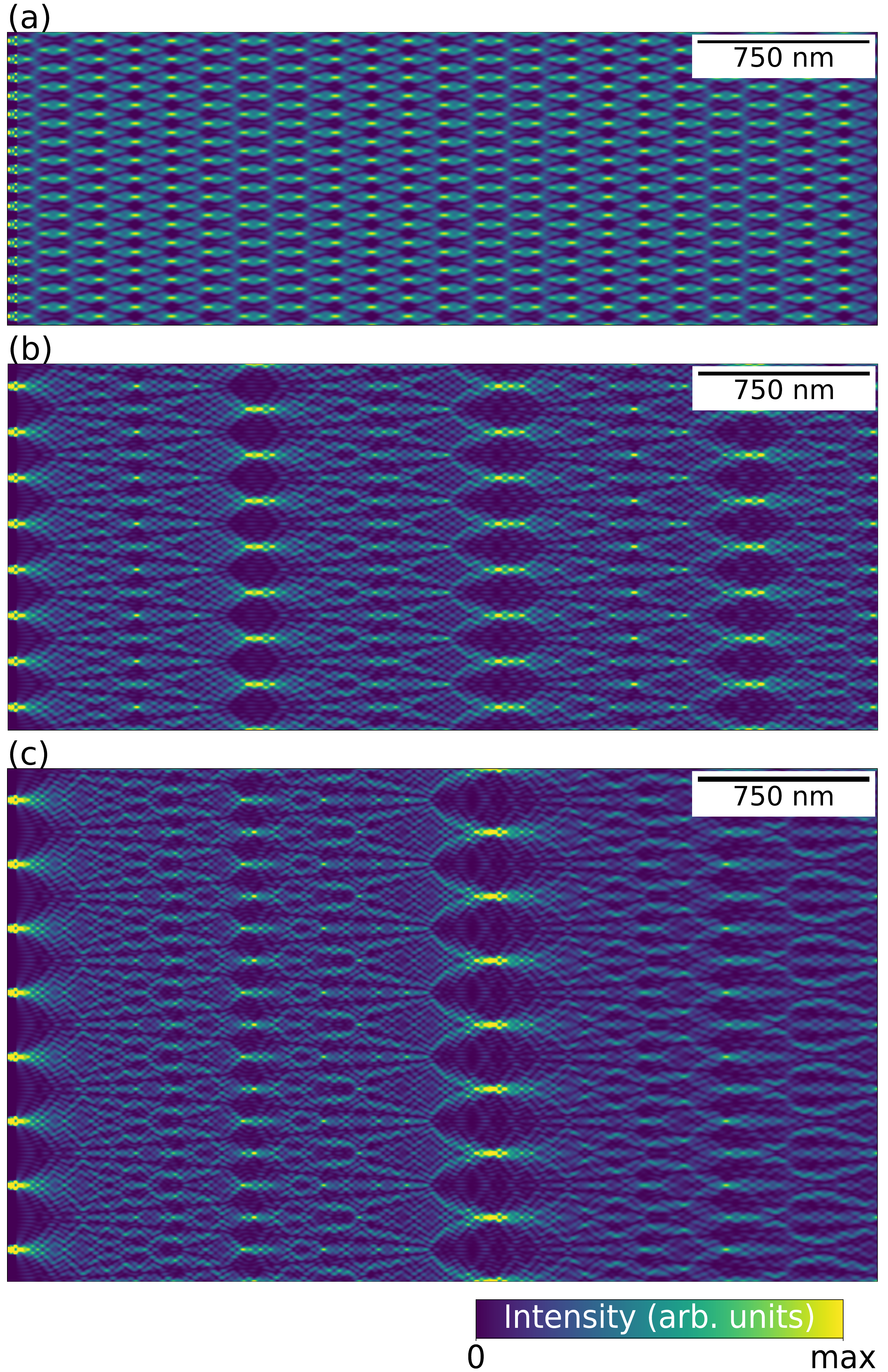}
\caption{Spin-wave intensity distribution at frequency of 40 GHz after passing through a one-dimensional diffraction grating with period of (a) 80 nm, (b) 200 nm, and (c) 280 nm. The grating is located at the left edge of each plot.}
\label{Fig:c40}
\end{center}
\end{figure}

\begin{figure}[htp]
\begin{center}
\includegraphics[width=\linewidth]{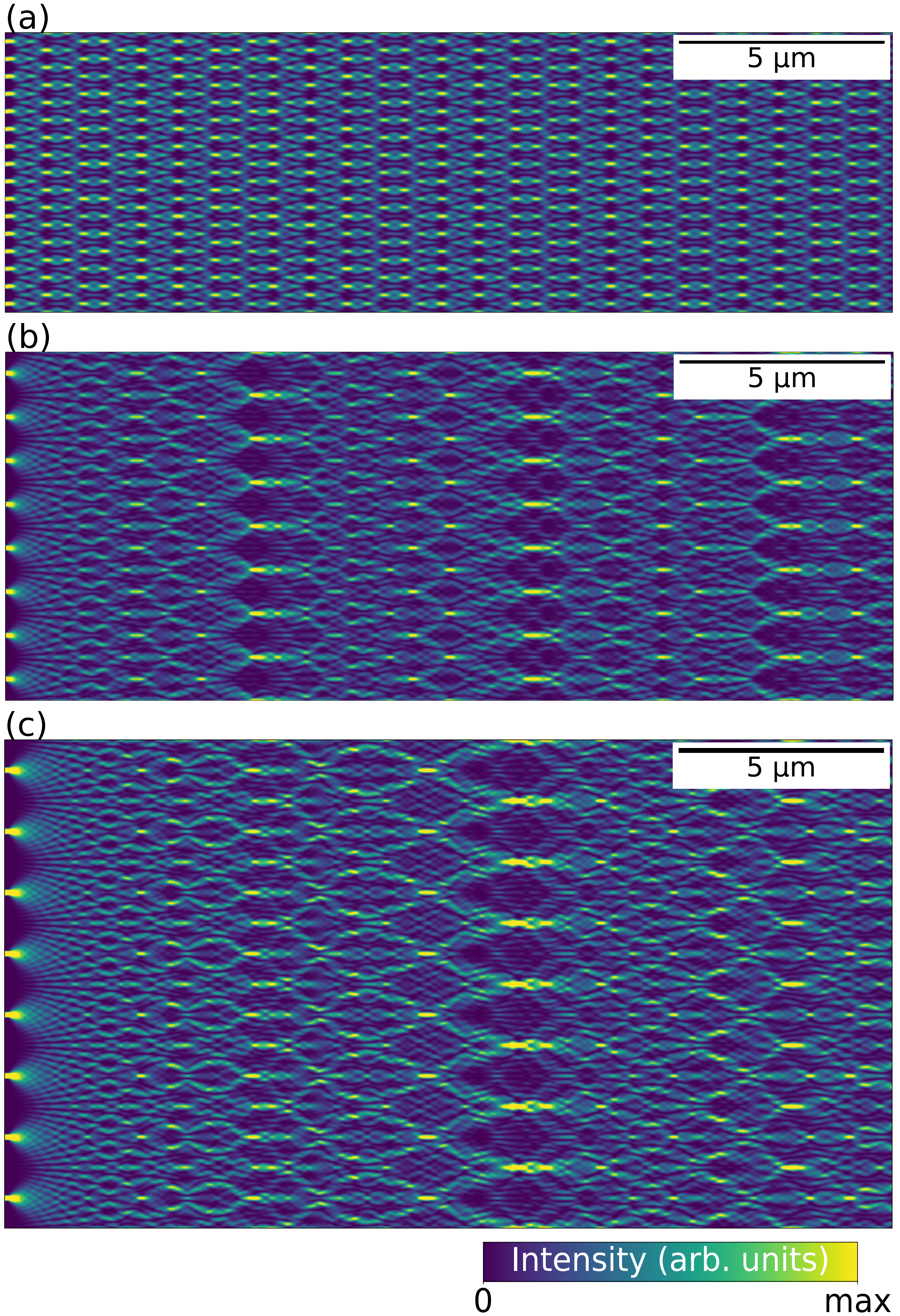}
\caption{Spin-wave intensity distribution field at frequency of 3 GHz after passing through a one-dimensional diffraction grating with period of (a) 400 nm, (b) 1000 nm, and (c) 1400 nm. The grating is located at the left edge of each plot.}
\label{Fig:c6}
\end{center}
\end{figure}

The simulated Talbot carpets are presented as the SW intensity maps, \textit{i.e.}, averaged in time, squared dynamic in-plane component of the magnetization,  $\left<m_x^2\right>$. These diffraction fields for SWs of frequencies 40 GHz and 3 GHz and for different values of $d$ are shown in Fig.~\ref{Fig:c40} and Fig.~\ref{Fig:c6}, respectively. To make comparison of interference images for the both frequencies more direct, every periods of the analyzed diffraction gratings were selected, so that for each pair of the results plotted in Figs.~\ref{Fig:c40} and \ref{Fig:c6} same kind of scaling takes place. In other words, visually the same (or very similar) carpets can be obtained for different choices of period and frequency. For more clarity, compare the peaks in the reciprocal (wavevector) space, see Fig.~\ref{Fig:kxky}. They approximately correspond to the same angle for 40 GHz and 3 GHz, while geometrical parameters were selected depending on frequency. 

Figure \ref{Fig:c40} shows the well-resolved Talbot carpets for SWs excited at frequency 40 GHz. It is observed that with the increasing period of the grating the Talbot length increases as well, and the effect of self-imaging is well-recognizable in the form of bright focal points. The resulting diffraction field obtained for 3-GHz-SWs (shown in Fig.~\ref{Fig:c6}) presents similar Talbot carpets. However, Talbot carpets for 3 GHz are slightly less regular than in the case of 40 GHz. Later in this paper, we will prove that the both Talbot carpets for SWs are consistent with theoretical predictions, in spite of the above-mentioned irregularity.

\begin{figure}[htp]
\begin{center}
\includegraphics[width=\linewidth]{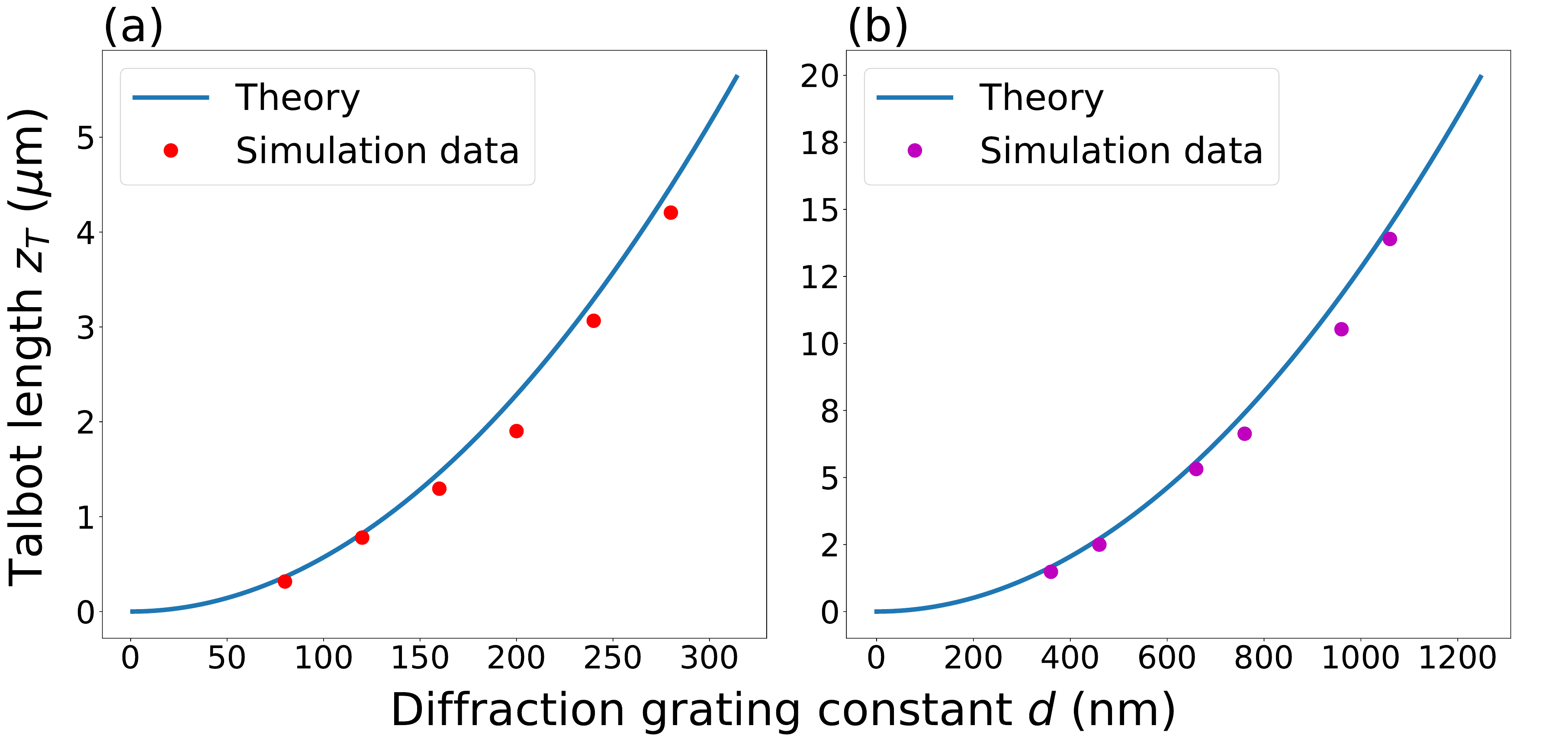}
\caption{Functions representing the Talbot length depending on the diffraction grating period from Eq.~(\ref{Eq:zt}) together with the data taken from micromagnetic simulations for (a) 40 GHz and (b) 3 GHz.}
\label{Fig:leng}
\end{center}
\end{figure}

The first, naturally occurring compliance test for the obtained numerical results is to compare the distance between the primary self-images (Fig. \ref{Fig:Sch_T}), Talbot length, with the theoretical predictions. For this purpose, we use Eq.~(\ref{Eq:zt}) as the function of $d$ for the two selected frequencies. As can be seen in Fig.~\ref{Fig:leng}, the simulation results well coincide with the general theory given in Sec. \ref{Sec:len}. Therefore, the simulated interference patterns behave very similarly to the Talbot carpets theoretically described in Sec.~\ref{Sec:TalC}. The Talbot lengths data used in Fig.~\ref{Fig:leng} come from the simulations carried out in an identical manner to those shown in Figs.~\ref{Fig:c40} and \ref{Fig:c6}. The distances between primary Talbot images were measured by analysing the lengths between individual subsequent intensity maximums (\textit{i.e.}, between each pair of integers, $m$ and $m+2$, see Eq.~\ref{Eq:zt} and Fig.~\ref{Fig:Sch_T}) on the two-dimensional Talbot carpets. The measured lengths were averaged to obtain one, possibly accurate value of Talbot length, for each of the selected values of $d$.

The obtained results show that the resulting SW Talbot carpets are in good agreement with the theoretical description of wave optics. The visualizations in Figs.~\ref{Fig:c40} and ~\ref{Fig:c6} show the distribution of SW intensity in the diffraction field according to Eq.~(\ref{Eq:psixz0}), which has been proven by comparing the numerically obtained and theoretically predicted Talbot lengths in  Fig.~\ref{Fig:leng}.

\subsection{Talbot carpets in wavevector space}

The next stage of the results analysis is to apply Fast Fourier Transform (FFT) to the obtained Talbot carpet visualizations, in order to transform the results from the space domain to the wavevector domain.

\begin{figure}[htp]
\begin{center}
\includegraphics[width=\linewidth]{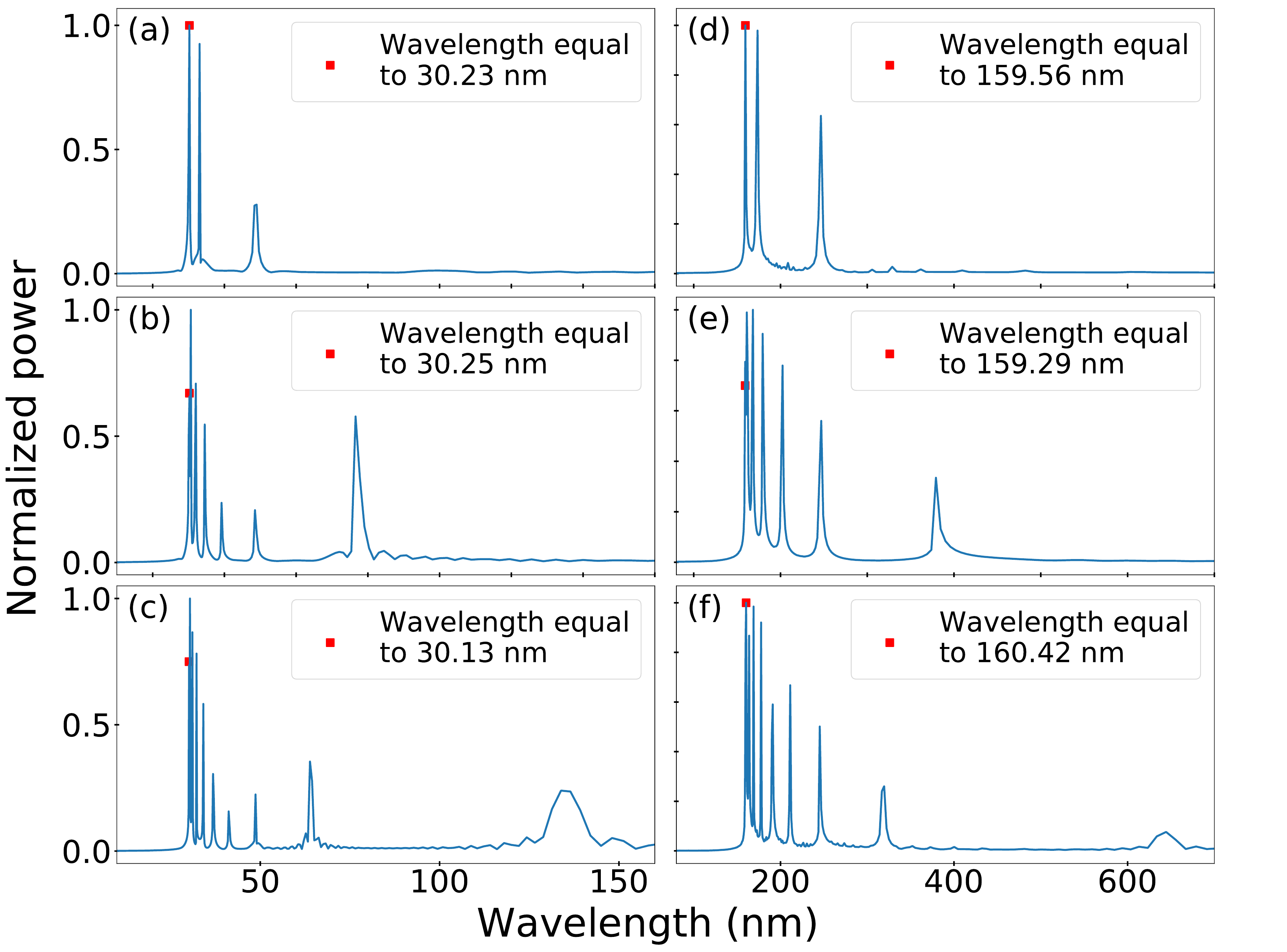}
\caption{SW spectra with marked first peaks responsible for the propagation direction wavelength. On the left panel, there are spectra from simulations performed for 40 GHz for period, $d$, which is equal to  (a) 80 nm, (b) 200 nm, and (c) 280 nm. On the right panel, we have analogous results for 3 GHz for period of (d) 400 nm, (e) 1000 nm, and (f) 1400 nm.}
\label{Fig:fft}
\end{center}
\end{figure}

\noindent To maintain greater accuracy, the absolute amplitude values of the FFT results (along the $z$-axis) were averaged over a perpendicular axis. The purpose of this averaging was to confirm that the frequency set used in the simulations corresponds to the theoretical values of the wavelengths given by the analytical dispersion relation \cite{kalinikos1986theory} and observe the discrete components of perpendicular wavevectors (\textit{i.e.}, transverse modes). In Fig.~\ref{Fig:fft}, we can see that for frequency of 40 GHz, the wavelength is about 30 nm, while for 3 GHz its value oscillates around 160 nm. These values satisfactorily coincide with the ones obtained from the dispersion relation, see Appendix \ref{Sec:App_dispersion}. Attention should also be paid here to the peaks occurring at larger wavelengths: they are associated with the transition of the plane SW through the periodic object. It is clear that the larger the value of $d$ is, the more pronounced the transverse components of the wavevector in our Talbot carpets are. A plane wave after encountering an obstacle (in our case -- in the form of a hole array) gains a non-zero component of the wavevector $k_x$ perpendicular to its propagation direction. Due to the nature of the obstacle, it also has discrete values being a multiple of $2 \pi/d$, see Section \ref{Sec:TalC} for details.\\

\begin{figure}[htp]
\begin{center}
\includegraphics[width=\linewidth]{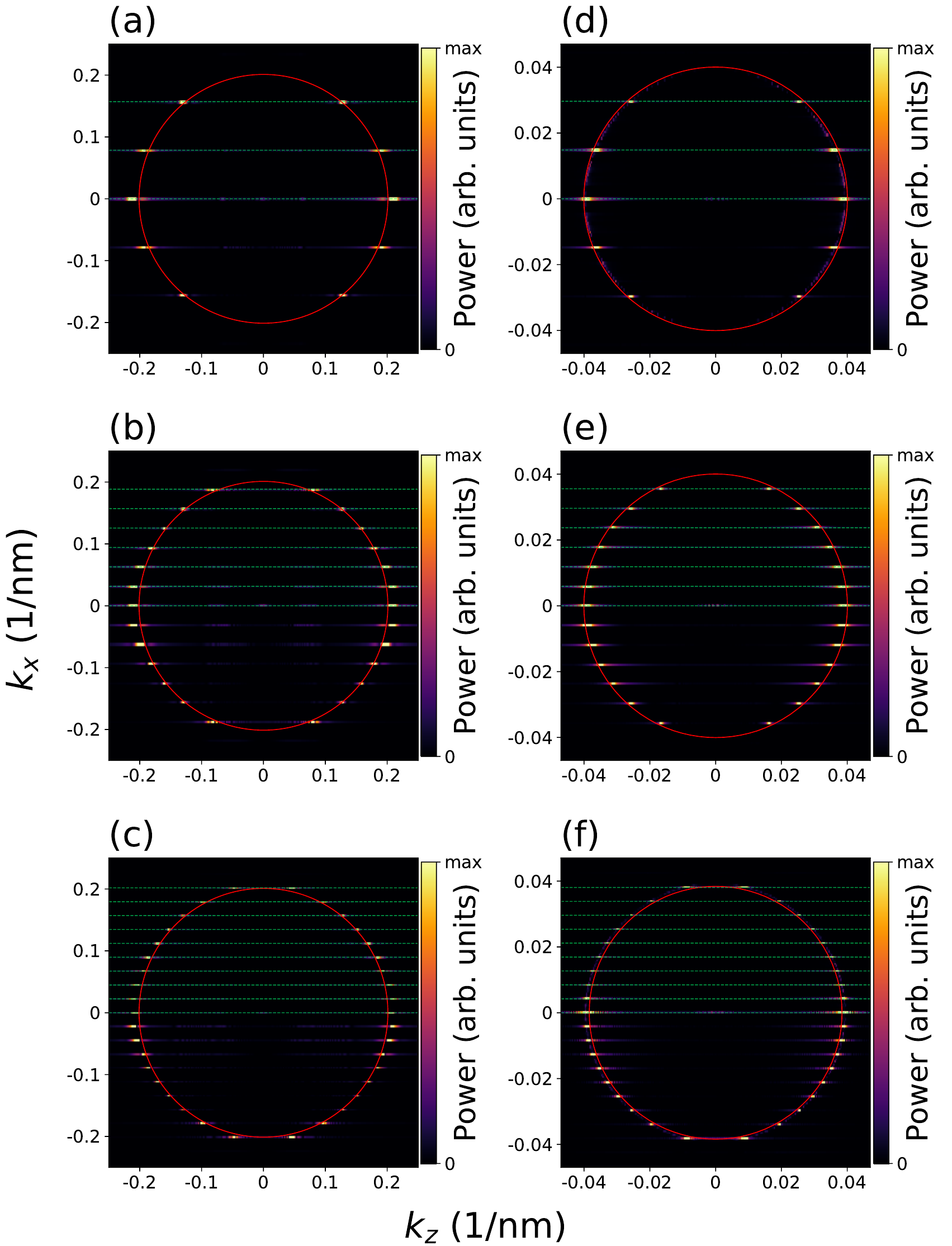}
\caption{Reciprocal-space maps of SW amplitude distribution at 40 GHz for the grating period equal to (a) 80 nm, (b) 200 nm, and (c) 280 nm, and at 3 GHz for the period equal to (d) 400 nm, (e) 1000 nm, and (f) 1400 nm. Isofrequency contours are presented as red circles. The quantized wavevector lines are shown in green.}
\label{Fig:kxky}
\end{center}
\end{figure}

To find transverse components of the wavevector co-creating the Talbot carpets, we perform two-dimensional FFT. In Fig.~\ref{Fig:kxky}, we can see the wavevector-space maps of SW amplitude, on the left side for 40 GHz and on the right side for 3 GHz, along with the marked isofrequency contours of the dispersion relation lines (IFDRLs) for the given frequency, and the lines specific for the quantized wavevector $ k_d $ (located on the $k_x$-axis). As we can read from these figures, the results obtained from the simulations agree very well with the theoretical data. Indeed, the individual peaks in the $(k_z,k_x)$ planes perfectly match the straight lines corresponding to the multiples of $ 2\pi/d $ provided in Eq.~(\ref{Eq:psixz0}). They are also located directly on the circular isofrequency lines, which confirms that the simulated propagation of SWs in a thin Py film is isotropic.\\
The results directly show that the Talbot effect can be successfully developed with SWs for its subsequent use for applications. The outcome visible in Fig.~\ref{Fig:kxky} also indicates that by manipulating the reciprocal space of the diffraction image we can predict its properties and shape in the real space. This feature is particularly useful, because it opens a way for the use of the Talbot effect for SWs in the systems, in which the location of self-images may provide information about the input signal or the nature of the previously encountered obstacle.

\section{Experimental feasibility}
\label{Sec:exp_feas}
\begin{figure}[htp]
\begin{center}
\includegraphics[width=\linewidth]{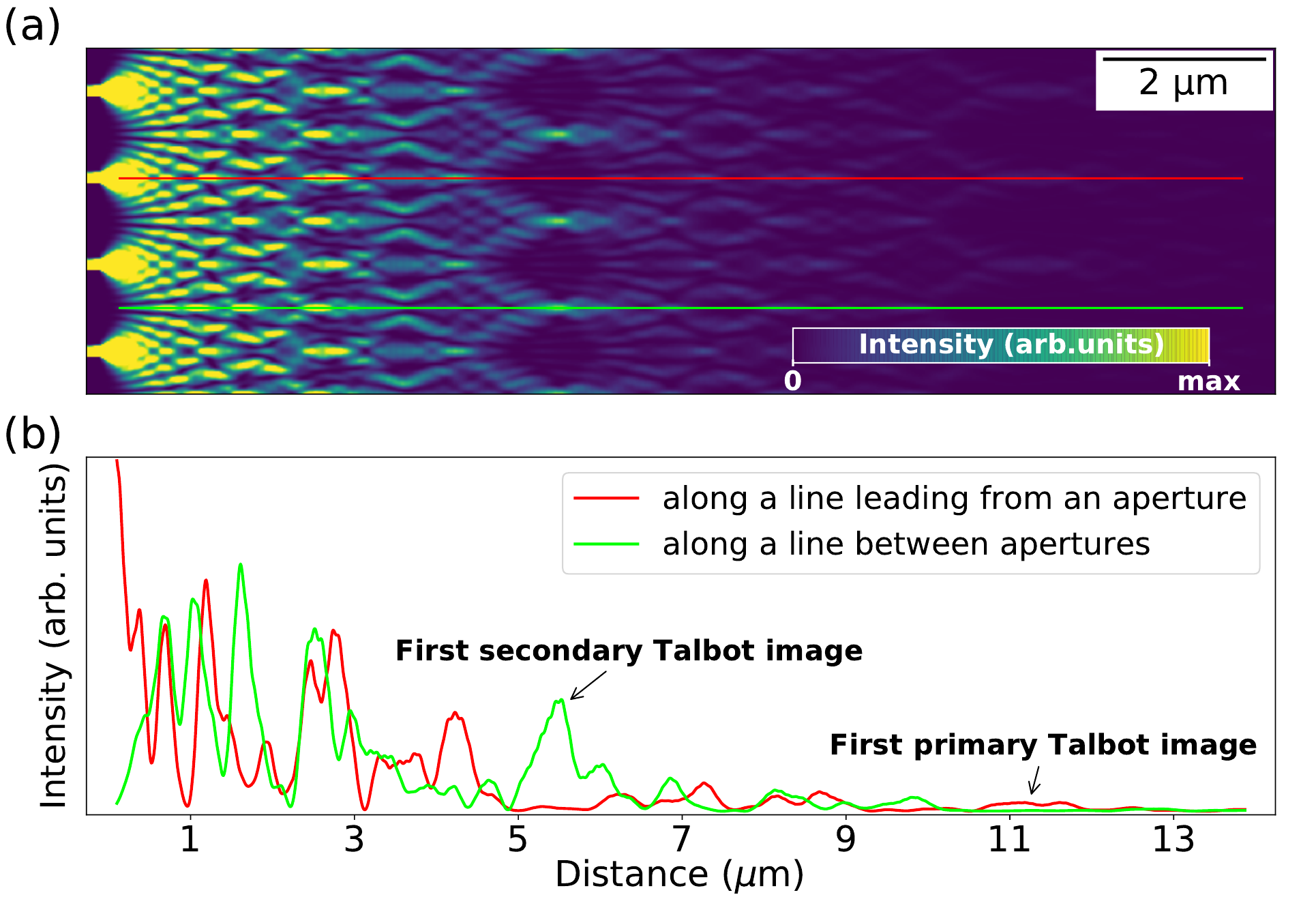}
\caption{Spin-wave Talbot effect in thin Py film with damping constant $\alpha = 0.005$ for 3 GHz and diffraction grating period $d=1000$ nm as (a) 2D intensity plot with mapped two lines (red and green) along which (b) 1D intensity plots were created. Diffraction grating is located at the left edge and the line colors are analogous on both graphs.}
\label{Fig:damp}
\end{center}
\end{figure}

The purpose of this work was, among others, to demonstrate the Talbot effect for SWs and to test its compliance with theoretical predictions derived from standard wave optics. For this reason, the system used for micromagnetic simulations assumed a very small damping constant ($\alpha = 0.0001$) as a factor not significant from the viewpoint of the SW Talbot effect demonstration. As we have successfully shown and described the effect, the next step was to test it in terms of real application, so we decided to increase the damping constant, as an inevitable parameter in magnonics, to a value of $\alpha = 0.005$, being characteristic for a thin Py film \cite{damping}. The simulation results are shown in Fig.~\ref{Fig:damp}.

Based on the simulation results for systems with a higher damping constant, we can see that the Talbot effect is clearly visible until the first secondary self-image is formed, so the diffraction field effective range, for this specific set of parameters, can be estimated by calculating the Talbot length formula (\ref{Eq:zt}). It is worth emphasizing that the SW intensity level on the first secondary self-image should be measurable using micro-focus Brillouin light scattering (micro-BLS) \cite{Brillouin, Brillouin2}. The simulation carried out shows that this effect, although significantly limited, can be observed in systems with higher damping, and its effectiveness will strongly depend on the choice of material, its dimensions and geometry of the diffraction grating.\\

\section{In-plane magnetization}
\label{Sec:in_plane}

\begin{figure}[htp]
\begin{center}
\includegraphics[width=\linewidth]{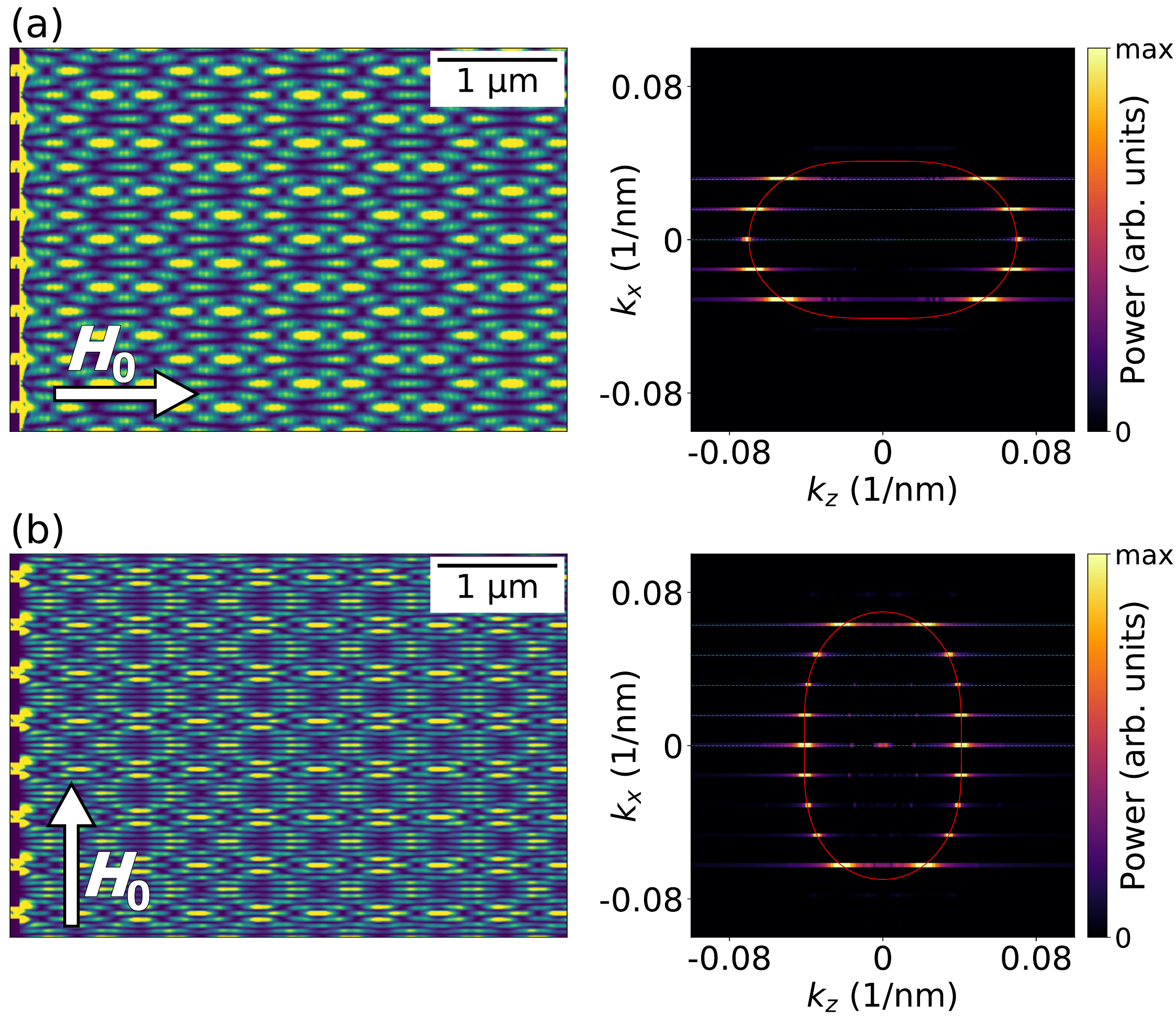}
\caption{Spin-wave intensity distribution fields after passing through a diffraction grating $d=500$ nm with corresponding representations in reciprocal spaces at frequency of 15 GHz for (a) BV and (b) DE geometry. Analytical isofrequency contours are presented as red ellipses and quantized wavevector lines separated by $\delta k_x=2\pi/d$ are shown in green.}
\label{Fig:bv_de}
\end{center}
\end{figure}

The demonstration of the Talbot effect for SWs was carried out so far for out-of-plane magnetization due to the isotropic properties of SW dynamics in this configuration -- thus the closest to electromagnetic waves in optically homogeneous medium. The proof of this can be seen directly in Fig. 7 in the form of circular IFDRLs matching with simulations data. This situation changes if the applied external magnetic field saturating magnetization is directed parallel (or non-perpendicular) to the plane of the film -- then we will observe, especially in the regime where dipolar interactions play significant role, an anisotropy in SW propagation and related caustic effects \cite{Veerakumar_Camley, Gieniusz_Ulrichs}. We have performed simulations of the Talbot effect for SWs in Damon-Eshbach (DE) and the backward volume (BV) magnetostatic wave geometry, where propagation of SWs is perpendicular and parallel, respectively, to the direction of external magnetic field of 0.1~T. The results are shown in Fig.~\ref{Fig:bv_de}, for the frequency equal to 15 GHz, for which the system anisotropy is already clearly visible in these magnetization configurations. The material parameters remain unchanged compared to the simulations in the Section \ref{Sec:Mic}.

The simulations show that it is possible to obtain SWs self-imaging also in much lower field than required in out-of-plane cases. This property can be decisive when it comes to the use of magnonic systems based on the Talbot effect in real devices. We can see, that Talbot carpets obtained for exactly the same material parameters differ significantly only under the change of external magnetic field configuration. The anisotropy of SW dynamics brings additional challenges, but it also extends the possibilities of controlling and manipulating SWs through an additional degree of freedom, such as an external magnetic field orientation. This opens the way for further research, \textit{e.g.}, on magnonic logic devices that can perform more than one function by changing the angle or the value of the external magnetic field, thus changing the distribution of self-images, which in turn will result in a different signal at the output. This is a clear benefit of using the Talbot effect for SWs, and is in contrast to analogous devices in photonics and electronics that cannot be programmed in this way.

The analysis of anisotropic effects associated with the use of lower frequencies in DE and BV geometries, in combination with counteracting of damping impact \cite{gieniusz2017switching}, provides a very interesting issue for further work of the Talbot effect in magnonics.

\section{Conclusions}
\label{Sec:conclusions}
In this paper, we have shown by using micromagnetic simulations, based on solutions of Landau-Lifshitz equation, that the Talbot effect occurs when SWs propagate in a thin ferromagnetic film, after passing through a periodic diffraction grating created in the film. We demonstrated that the properties of SW self-imaging are consistent with the theoretical predictions based on the general formalism of wave optics. Thus, it can be used to describe this phenomenon quite accurately in the considered range of parameter variation. This compliance cannot be introduced in advance, \textit{i.e.}, based on the knowledge on the Talbot effect in optics that is described by Maxwell equations. Rather, Landau-Lifshitz nonlinear equation describing SW propagation must be solved for this purpose. This has systematically been done in our study. By performing micromagnetic simulations in Py film with characteristic damping, we showed that the observation of the first secondary Talbot images shall be feasible with standard micro-BLS. We expect, that in Yttrium Iron Garnet thin films, the (first) primary images can be reached due to ten times smaller damping \cite{yttrium}. Moreover, we demonstrated that the Talbot effect exists for SWs in the out-of-plane magnetized and in-plane magnetized film when the SW isofrequency contours of the dispersion relation are isotropic and anisotropic, respectively.

The obtained results open an avenue to practical application of the Talbot effect in future magnonic devices. Indeed, it is a promising phenomenon from the viewpoint of analysis, control, and manipulation of SW propagation. That is why it may find applications in magnonics, where devices of such a type could be used in signal processing, \textit{e.g.}, in logic circuits and SW analyzers. In the coming years, further theoretical research, experimental demonstration, and development of prototypes of magnonic devices based on the SW Talbot effect are expected to occur. This work is the first step in understanding the main features and assessing the potential of the studied effects.

\begin{acknowledgments}
The research leading to these results has received funding from the National Science Centre of Poland, Project No. UMO-2015/17/B/ST3/00118. The simulations were partially performed at the Poznan Supercomputing and Networking Center (Grant No.~398). P.G. acknowledges support from the  National Science Centre of Poland under OPUS funding, Project No. UMO-2019/33/B/ST5/02013.
\end{acknowledgments}

\appendix

\section{Derivation of the Talbot length -- model analysis I}
\label{Sec:App_A}

In order to determine the Talbot length, we analyze the intensity (distribution of a flux density) as a function of a distance between the grating and a given observation point $P$ in the diffraction field, according to Refs. \cite{hiedeman-9, winthrop-10, montgomery-11}.

\begin{figure}[htp]
\begin{center}
\includegraphics[width=\linewidth]{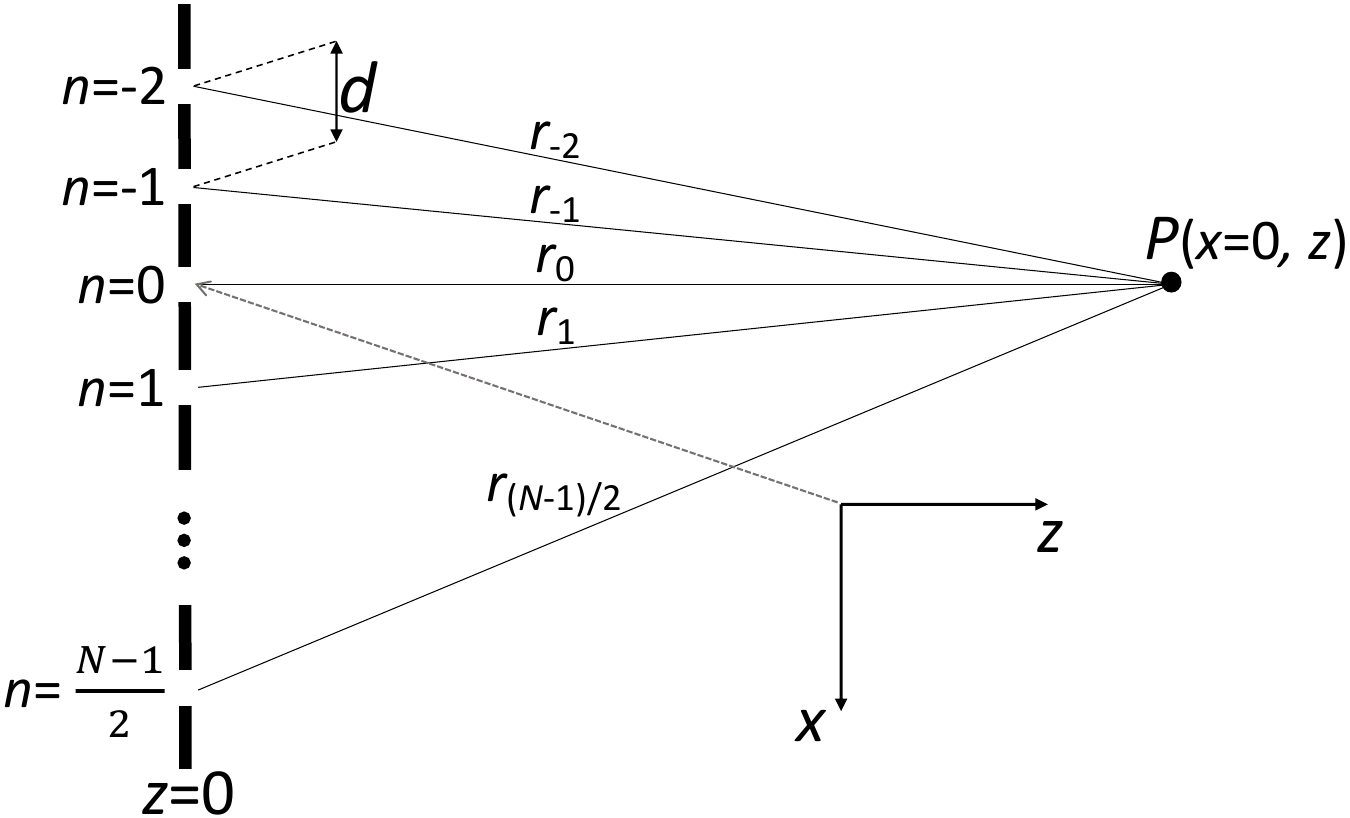}
\caption{Diagram showing the situation described in this section together with relevant markings -- a diffraction grating consisting of $N$ sources distant from the analyzed point $P$ by $r_n$, where $n$ is a number indicating the given aperture. The center of coordinate system is in the aperture $n=0$.}
\label{Fig:superpos}
\end{center}
\end{figure}

\noindent At point $P$ away from the source (being one of the apertures -- Fig. \ref{Fig:superpos}) by distance $r_n$, the resulting wave $\mathit{\Psi}(\vec{r},t)$ can be written as a superposition of the $N$ waves in the following form:
\begin{equation}
\mathit{\Psi}(r,t)=\Re\left(\sum_{n=-(N-1)/2}^{(N-1)/2}\!\!\!\!\!\!C_{n}e^{-i\omega{t}}[\cos{(kr_n)}+i\sin{(kr_n)}]\right).
\label{Eq:superpos}
\end{equation}
Then, intensity can be defined as follows:
\begin{equation}
I(r)=\langle{\:\mathit{\Psi}^2(r,t)\:}\rangle{_t} 
= \frac{(NC)^2}{2}\left(\sum_{n=-(N-1)/2}^{(N-1)/2}\!\!\!\cos{(k r_n)}\right)^2,
\label{Eq:int}
\end{equation}
where
\begin{equation}
r_n=\sqrt{z^2+(nd+x)^2}.\label{Eq:rn}
\end{equation}
In order to simplify further analysis, we neglect the constant $(NC)^2/2$, remembering that $N$ is the number of apertures of the diffraction grating, $C$ is a numerical factor, and we consider the calculations for the point $P$ such that $x = 0$ (as shown in Figure \ref{Fig:superpos}). This results in the distribution of intensity in the diffraction field being a function of $z$ coordinate:
\begin{equation}
I(z,x=0)=\left(\sum_{n=-(N-1)/2}^{(N-1)/2}\cos(\;{\frac{2\pi}{\lambda}\sqrt{z^2+(nd)^2}})\right)^2. \label{Eq:Izx0}
\end{equation}
To examine the course of the expression (\ref{Eq:Izx0}), the following substitution has been made:
\begin{equation}
U_n=\frac{2\pi}{\lambda}\sqrt{z^2+(nd)^2}. \label{Eq:Un}
\end{equation} 
As we are looking for the maximum of \ref{Eq:Izx0} expression in the $z$ function (self-image of the source intensity), the $\cos^2{(U_n)}$ has to take the value $1$ (for any $n$). We can easily see that it takes place only when  $U_n=\pi{b}$, for $b$ being a natural number ($b =$ $0$, $1$, $2$, $\ldots$). 
To extract the classical formula for the Talbot length, now we assume that 
\begin{equation}
\left(\frac{nd}{z}\right)^2 \ll 1
\end{equation}
and use Taylor's expansion as follow:
\begin{equation}
\sqrt{1+\left(\frac{nd}{z}\right)^2} \approx  1+\frac{1}{2}\left(\frac{nd}{z}\right)^2. \label{Eq:Tay}
\end{equation}
All terms of the Taylor series above the second one were omitted. We can now substitute the approximate result of (\ref{Eq:Tay}) to (\ref{Eq:Un}), and remembering that $U_n=\pi{b}$, we get the condition
\begin{equation}
\frac{2z}{\lambda}+\frac{(nd)^2}{\lambda{z}}=b. 
\end{equation}
Note, that the fraction $2z/\lambda$ is, in fact, a certain integer, indicating the doubled number of wavelengths, $\lambda$, along the $z$-axis. We can therefore substitute $2z/\lambda = b_0$, thanks to which we get 
\begin{equation}
z=\frac{n^2}{b-b_0} \frac{d^2}{\lambda}.
\label{Eq:z}
\end{equation}
Since for all $n$ there exists an integer $b-b_0$, that satisfies Eq.~(\ref{Eq:z}), every aperture (\textit{i.e.}, every circular wave source) contributes to the resulting phase at a certain position, which can be finally found by using the following formula for the Talbot length:
\begin{equation}
z_T=m\frac{d^2}{\lambda}, \label{Eq:ztt}
\end{equation}
where $m$ is an integer specifying the number of subsequent self-images.

\section{Derivation of the Talbot length -- model analysis II}\label{Sec:App_B}

\begin{figure}[htp]
\begin{center}
\includegraphics[width=\linewidth]{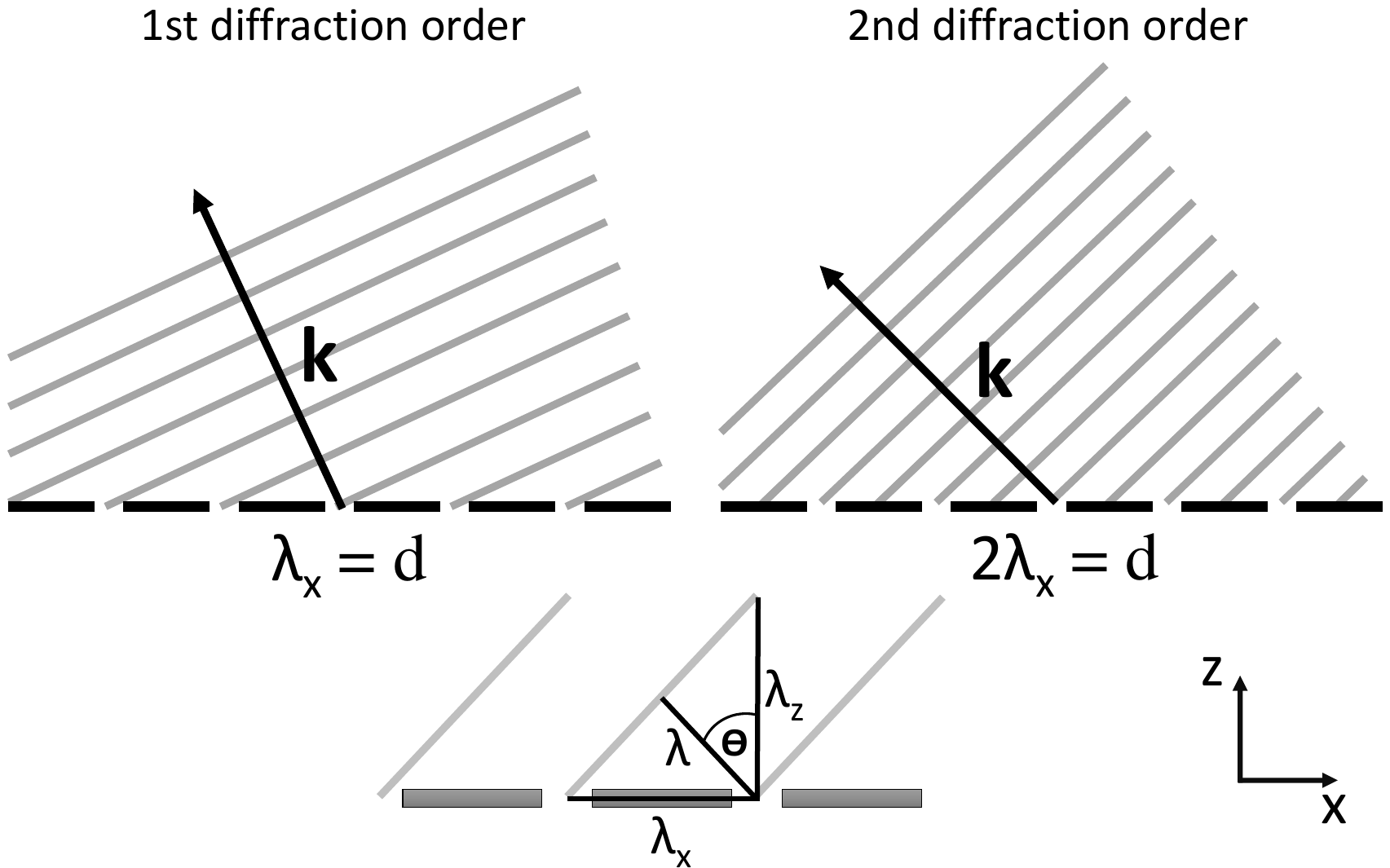}
\end{center}
\vspace{-0.6cm}
\caption{Propagation scheme of two diffraction orders of the plane wavefronts behind a diffraction grating, in accordance with the rule described by Eq.~(\ref{Eq:bd}). Gray lines are the wavefronts with equal phases, while the black arrows determine the directions of wavevector. A wavelength representations on the axes are marked below. \label{Fig:p2}} 
\end{figure}

Another, slightly more intuitive method of determining the Talbot length \cite{cowley, winthrop-10}, uses the fact that each order (greater than 0) of the wavefront, which has passed through the diffraction grating, must have, in the direction parallel to the apertures plane (\textit{i.e.}, along $x$-axis), the period that is a natural multiple of the distance between them:
\begin{equation}
b\lambda_x=d, \label{Eq:bd}
\end{equation}
where $b$ denotes natural numbers, $b = $ $0$, $1$, $2$, $\ldots$, and $\lambda_x$ is the distance between wavefronts along the $x$-axis ($x$-direction modulation for simplicity called ,,horizontal wavelength'' while $\lambda_z$ -- ,,vertical wavelength'', remembering that they are not wavelength projections).
The described physical situation is presented in Fig.~\ref{Fig:p2}.

In order to find the dependence between those particular wavelengths, the geometrical property presented in Fig.~\ref{Fig:p2} is used, from which a simple relationship can be obtained as follows:
\begin{equation}
\frac{1}{\lambda^2}=\frac{1}{\lambda_x^2}+\frac{1}{\lambda_z^2}.
\label{Eq:lambda_geom}
\end{equation}
Using the above introduced Eq.~(\ref{Eq:bd}) and simple mathematical derivations, we get the following formula for the ,,vertical wavelength'':
\begin{equation}
\lambda_z=\frac{\lambda}{\sqrt{1-(\frac{\lambda{b}}{d})^2}}. \label{Eq:lambda_z}
\end{equation}
As we know, a period of repetitive intensity modulation in the direction parallel to the $z$-axis, after passing through the periodic structure, has a strictly defined value. Therefore we considered points on this axis, for which the secondary waves coming from each of the apertures will have the same phase -- it is the condition of occurrence of the constructive interference. The phase distribution of the waves along the $z$-axis is introduced as
\begin{equation}
\phi_z=k_zz=\frac{2\pi}{\lambda_z}z. \label{Eq:phiz}
\end{equation}
For the points on the $z$-axis that are a multiple of the wavelength $\lambda_z$, the phase will be constant and equal to $2\pi$. By combining Eq.~(\ref{Eq:lambda_z}) and Eq.~(\ref{Eq:phiz}), we obtain the phase distribution of the diffraction image along the $z$-axis:
\begin{equation}
\phi_z=\frac{2\pi{z}}{\lambda}\sqrt{1-\left(\frac{\lambda{n}}{d}\right)^2}. \label{Eq:phiz2}
\end{equation}
Assuming that the grating constant is much larger than the wavelength, it can be written: 
\begin{equation}
\left(\frac{\lambda{n}}{d}\right)^2\ll 1. 
\end{equation}
Next, again using Taylor's expansion, we obtain:
\begin{equation}
\phi_z \approx \frac{2\pi{z}}{\lambda}\left(1 -\frac{1}{2}(\frac{\lambda{n}}{d})^2\right)=\frac{2\pi{z}}{\lambda}-\frac{\pi{n^2}\lambda{z}}{d^2}. \label{Eq:phiz3}
\end{equation}

Having already expressed the phase distribution in the diffraction field along the $z$-axis, we could think about particular points we are interested in. As we know, the Talbot length determines the distance between the successive maxima of intensity perpendicular to the diffraction grating plane and coming from one of the apertures. What is obvious, the maximum intensity also occurs on the apertures themselves (at $x = 0$). Therefore, their subsequent repetitive modulation will be the reproduction of the image from the beginning of the system. Thus, the reconstruction of the diffraction grating itself (self-imaging) takes place. We are looking for a constructive  interference along the $z$-axis for each of the values $x = nd$, which correspond to the subsequent apertures positions. The first term in the right-hand side of Eq.~(\ref{Eq:phiz3}) does not depend on which aperture we choose as a reference point, so we should make its second term a multiple of the full period. Then, this formula could fulfil the condition of constructive interference, so we can write in our case
\begin{equation}
\phi_z=\frac{2\pi{z}}{\lambda}-2\pi{n^2}. \label{Eq:phiz4}
\end{equation}
Comparing Eq.~(\ref{Eq:phiz3}) with the above given condition of constructive interference, we can see that they are identical only if  ${\lambda{z}}/{2d^2}=1$, which leads us to the desired Talbot length, \textit{i.e.}, 
\begin{equation}
z_T=\frac{2d^2}{\lambda}. \label{Eq:z_T}
\end{equation}
In Eq.~(\ref{Eq:z_T}), the obtained Talbot length is multiplied by $2$, which  can be omitted in the general case. As we can see in Eq.~(\ref{Eq:zt}), the factor by which $d^2/\lambda$ is multiplied determines only the order, \textit{i.e.}, the number of the Talbot image analyzed in the sequence, and the fact whether the self-image is shifted laterally in a phase (secondary image, for an odd factor) regarding the original image or is it the phase compatible with that image  (primary image, for an even factor), see Fig.~\ref{Fig:Sch_T}. Equation (\ref{Eq:z_T}) gives the distance between the source and the first-order primary self-image.\\
The laterally shifted Talbot image would be, in turn, extinguished by taking over the condition for destructive interference 
\begin{equation}
\phi_z=\frac{2\pi{z}}{\lambda}-\pi{n^2},
\end{equation}
which yields 
\begin{equation}
z_T=\frac{d^2}{\lambda},\label{Eq:zT3}
\end{equation}
being the Talbot length for the odd factor $m =1$, according to Eq.~(\ref{Eq:zt}).

\section{Analysis of computational uncertainties}
It should be emphasized \cite{phase}, that the analytical formulas obtained in Appendix A and Appendix B give correct results in the systems, for which the approximations  
\begin{equation}
d\gg{\lambda} \label{Eq:d_lamb}
\end{equation}
and
\begin{equation}
z\gg{d}. \label{Eq:z_d}
\end{equation}
are applicable. Thanks to them, it was possible to apply Taylor's expansion in each of the cases described by Eqs.~(\ref{Eq:kztayl}, \ref{Eq:Tay}, and \ref{Eq:phiz3}). While the condition (\ref{Eq:z_d}) can be easily met in most cases, the condition (\ref{Eq:d_lamb}) is no longer so obvious. Indeed, for diffraction gratings whose spatial period is comparable to a given wavelength, then Eq.~(\ref{Eq:z_T}) can cause significant discrepancies as compared to experimental or numerical data.\\

\begin{figure}[htp]
\begin{center}
\includegraphics[width=\linewidth]{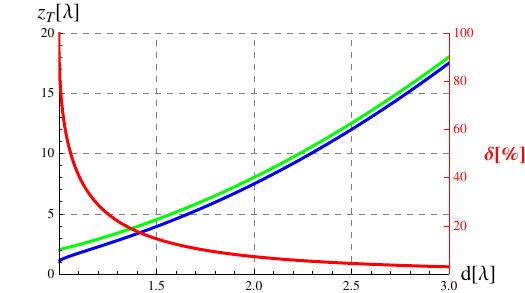}
\caption{Graph of the Talbot length as a function of the diffraction grating constant. General form and the form derived in this paper are shown by blue and green lines. The discrepancy function is shown by red line. \label{Fig:por}}
\end{center}
\end{figure}

In the case when $d \approx \lambda$, the verification of experimental and numerical results is based on the exact, general solution derived by Lord Rayleigh in 1881 \cite{Rayleigh81}, defining the Talbot length as 
\begin{equation}
z_{T_{\text{gen.}}}=\frac{\lambda}{1-\sqrt{1-(\frac{\lambda}{d})^2}}. \label{Eq:zt_R}
\end{equation}
To quantify the difference of results obtained from Eqs.~(\ref{Eq:z_T}) and (\ref{Eq:zt_R}), a graph of the function $z_T(d)$ generated from the both equations is presented in Fig.~\ref{Fig:por}.

\noindent The discrepancy function (expressed in per cent) is defined as the relative difference of values obtained from Eqs.~(\ref{Eq:zt_R}) and (\ref{Eq:z_T}):
\begin{equation}
\delta=\frac{|z_{T_{\text{gen.}}}-z_T|}{z_{T_{\text{gen.}}}}\cdot 100\%.
\label{Eq:error}
\end{equation}
Analyzing the results presented in Fig.~\ref{Fig:por}, we can conclude that the discrepancy function increases as the diffraction grating period decreases, and reaches $100\%$ when $d = \lambda$. Thus, it is justified to use Eq.~(\ref{Eq:z_T}) only in the systems, where the grating constant is properly greater than the incident wave length. Otherwise, it is necessary to use its general form (\ref{Eq:zt_R}).

\section{Simulations details}\label{Sec:App_Simulation}
The performed simulations consist of two parts. Firstly, the static magnetic configuration has been reached which, in turn, was perturbed by a source emitting plane SWs incident normally on the grating. SWs have been excited by the continuously applied RF field, localized on the left side of the diffraction grating within a 15-nm-wide region. SWs have been continuously induced until reaching the resultant steady state, \textit{i.e.}, the state when fully evolved interference pattern is formed. Along the film's edges perpendicular to the grating, the periodic boundary conditions have been introduced to mimic an infinitely long diffraction grating. In contrast, at the edges parallel to the grating, the absorbing boundary conditions with a gradually increasing damping constant have been used \cite{venkat2018absorbing}. The out-of-plane simulations have been performed for two frequencies: 40 GHz and 3~GHz which correspond to the wavelengths 31.2 nm and 156.9 nm, respectively (see Appendix~\ref{Sec:App_dispersion}). Pure exchange SWs are definitely easier to model for higher frequencies, since the wavelength for 40 GHz is only a few times greater than the exchange length ($l_\mathrm{ex}=\sqrt{2A_\mathrm{ex}/(\mu_0M_\mathrm{S}^2)}=6$ nm). However, this case is not yet accessible experimentally. Therefore, we have decided to perform simulations for a more realistic regime to check whether this effect is obtainable for frequencies and wavelengths available in contemporary laboratories, which would directly affects its application potential. We also investigated the spin-wave Talbot effect in BV and DE configurations for 15 GHz, where the isofrequency contours indicate that the calculations were made for the dipole-exchange regime, \textit{i.e.}, both dipole and exchange interactions play a significant role. Much smaller required external magnetic field (in-plane eleven times smaller than out-of-plane in our simulations) is a step towards experimental prototypes.\\
For each analyzed SWs frequency, the studied system was discretized by the $5\times 5\times 5$ $\mathrm{nm^3}$ unit cells.

\section{Spin wave dispersion}
\label{Sec:App_dispersion}

The analytical theory of SWs in thin ferromagnetic films was developed by Kalinikos and Slavin in Ref.~\cite{kalinikos1986theory}. Following that theory, in the linear approximation the dispersion relation, where wavevector $k$ propagates in the film plane at an angle $\varphi$ with respect to the direction of the external magnetic field $H_{0}$ projected onto the film plane, and the $H_{0}$ and static magnetization vector $M_{\mathrm{S}}$ form an angle $\vartheta$ with the normal to the film plane, takes the form:
\begin{equation}
\omega^{2}=\left(\omega_{\mathrm{H}}+l_{\mathrm{ex}}^{2}\omega_{\mathrm{M}}k^{2}\right)\left(\omega_{\mathrm{H}}+l_{\mathrm{ex}}^{2}\omega_{\mathrm{M}}k^{2}+\omega_{\mathrm{M}}F\left(\varphi,\vartheta\right)\right), 
\label{eq:dispersion}
\end{equation}
where $\omega=2\pi f$ is the angular frequency of SWs, $f$ is the frequency, $\mu_{0}$ is the vacuum permeability, $\omega_{\mathrm{H}}=\left|\gamma\right|\mu_{0}(H_{0}-M_{\mathrm{S}})$, $\omega_{\mathrm{M}}=\gamma\mu_{0}M_{\mathrm{S}}$, and the function $F(\varphi,\vartheta)$ is defined as:
\begin{multline}
F(\varphi,\vartheta)=P+\sin^2{(\vartheta)}\\\cdot\left[1-P\left(1+\cos^2{(\varphi)}+M_{\mathrm{S}}\frac{P(1-P)\sin^2{(\varphi)}}{H_0+l^2_{\mathrm{ex}}M_{\mathrm{S}}}\right)\right],
\label{eq:f_param}
\end{multline}
where
\begin{equation}
P=1-\frac{1-e^{-kL}}{kL},
\label{eq:P}
\end{equation}
and $L$ is the thickness of the analyzed ferromagnetic film. The contribution of dipolar interactions to the SWs dynamics is expressed by the term $F(\varphi,\vartheta)$ and the effect of the exchange interaction is represented in Eq.~(\ref{eq:dispersion}) by the terms proportional to $k^{2}$. 

\begin{figure}[htp]
\begin{center}
\includegraphics[width=\linewidth]{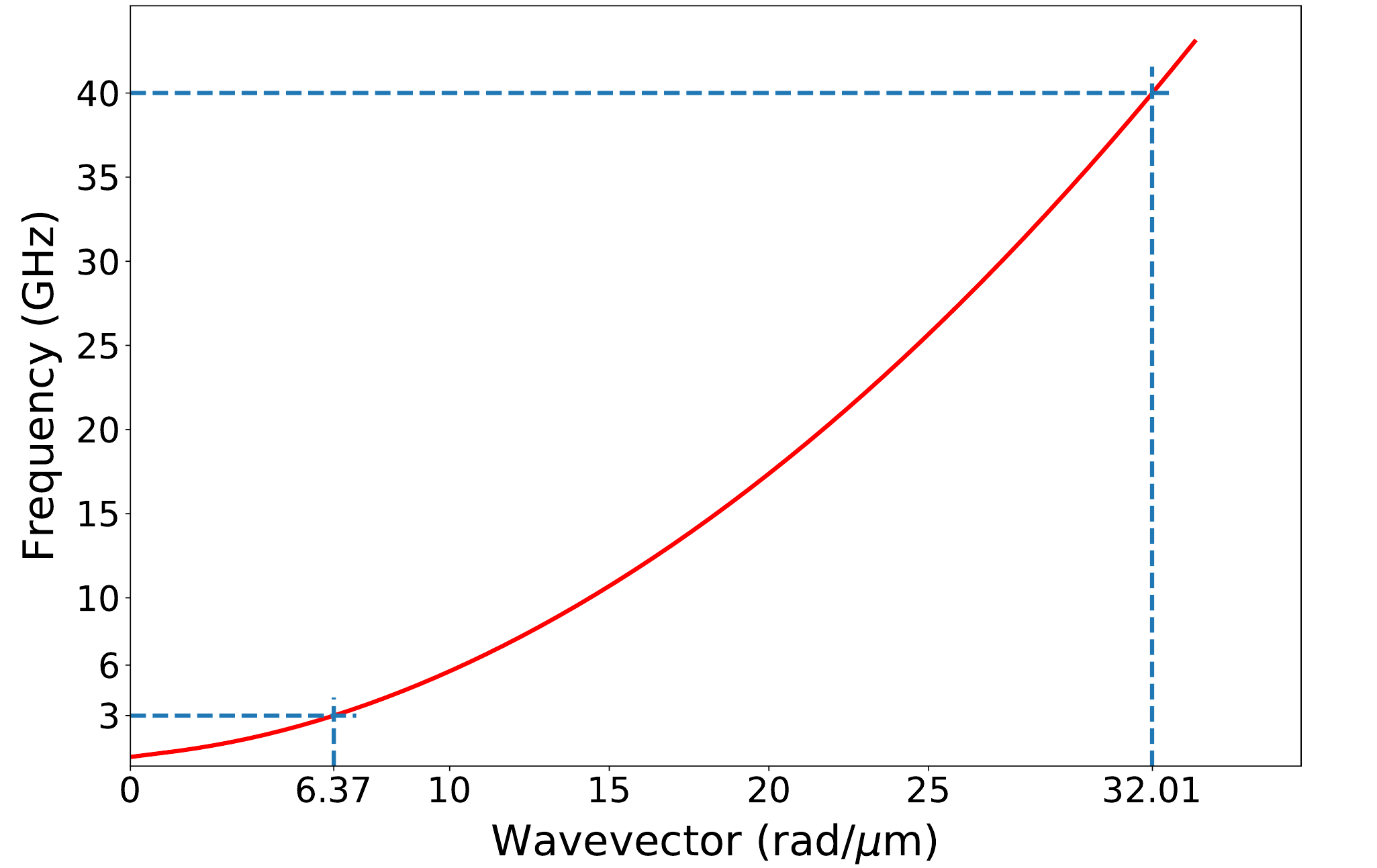}
\caption{The dispersion relation for 5-nm-thick Py film in the presence of 1.1 T out-of-plane magnetic field. The two marked values of the wavevector correspond to the frequencies 3 GHz and 40 GHz.}
\label{Fig:disp}
\end{center}
\end{figure}

The dispersion relation of the simulated Py film is presented in Fig.~\ref{Fig:disp}. We can see that for 40 GHz the wavevector value along the $z$-axis $ k_z $ is equal to $32.01$ $\mathrm{rad/\mu m}$, which corresponds to the wavelength $\lambda_z = 31.24 $ nm. For 3 GHz, $ k_z = 6.37$ $\mathrm{rad/\mu m}$, so $ \lambda_z = 156.92 $ nm. 

\bibliographystyle{apsrev4-1}
\bibliography{bibliography}

\end{document}